\documentstyle[preprint,aps]{revtex}
\begin{document}
\draft

\title{Spin-Flavour Oscillations and Neutrinos from SN1987A}

\author{M. Br\"uggen} \address{Institute of Astronomy, University of
Cambridge, Madingley Road, Cambridge CB3 0HA, UK} 
%\date{\today}
\maketitle
\begin{abstract}
The neutrino signal from SN1987A is analysed with respect to
spin-flavour oscillations between electron antineutrinos,
$\bar{\nu}_{e}$, and muon neutrinos, $\nu_{\mu}$, by means of a
maximum likelihood analysis.

Following Jegerlehner {\em et al.} best fit values for the total
energy released in neutrinos, $E_{\rm t}$, and the temperature of the
electron antineutrino, $T_{\bar{\nu}_{e}}$, for a range of mixing
parameters and progenitor models are calculated. In particular the
dependence of the inferred quantities on the metallicity of the
supernova is investigated and the uncertainties involved in using the
neutrino signal to determine the neutrino magnetic moment are pointed
out.

\end{abstract}
\pacs{14.60.St, 97.60.Bw}
%\pagebreak 
\section{Introduction}

At the end of their lives massive stars explode in a type II supernova
event releasing almost their entire binding energies in form of
neutrinos of all flavours. Depending on the model for the progenitor,
the equation of state and the treatment of the neutrino transport the
theoretical prediction for the total energy emitted in neutrinos is

\begin{equation}
E_{\rm t} \approx 2 - 4 \times 10^{53}{\rm erg}
\end{equation}

and for the mean neutrino temperature is

\begin{equation}
T_{\nu}\approx\left \{ \begin{array}{ll}
3-4\  {\rm MeV} & \text{for $\nu_{e}$},\\
5-6\  {\rm MeV} & \text{for $\bar{\nu}_{e}$},\\
7-9\  {\rm MeV} & \text{for all other species.}
\end{array}
\right .
\end{equation}
\cite{Bu,BuLa} The inferred values from the observed neutrino signal
from the explosion of SN1987A in the Large Magellanic Cloud (LMC) are
$E_{\rm t}\approx 2-3 \times 10^{53}{\rm erg}$ and
$T_{\bar{\nu}_{e}}\approx 3$ MeV \cite{Hi,Bi,Lo}. While the energy,
$E_{\rm t}$, is in good agreement with supernova theory, the
temperature of the electron anti-neutrinos lies below the predicted
range.  Neutrino oscillations could significantly alter the energy
spectra as observed by neutrino detectors on Earth and have
implications for the interpretation of the observed neutrino
signal. In fact an admixture of the hotter $\nu_{\mu}$-spectrum
through oscillations would shift the inferred value for
$T_{\bar{\nu}_{e}}$ towards lower values thus making it even more
difficult to reconcile theory with observation.

This analysis follows closely the method by Jegerlehner {\em et al.}
\cite{Je}, who investigated the implications of flavour oscillations
$(\bar{\nu}_{e}\leftrightarrow \bar{\nu}_{\mu})$ on the interpretation
of the supernova neutrino signal using a maximum likelihood
method. The impact of the MSW-effect on the neutrino spectra has also
been studied in Ref. \cite{At,Sm,Ke}.  For a `normal' mass hierarchy
(i.e. $m_{\nu_{e}}<m_{\nu_{\mu}}$) flavour oscillations between
antineutrinos are effective only for a very limited choice of mixing
parameters. However spin-flavour (SF) oscillations, which arise from
the interaction of the (hypothetical) magnetic moment of the neutrino
with an external magnetic field, could have a serious effect for a
wide range of reasonable parameters.

In this paper I will investigate the implications of SF-conversion
between right-handed $\bar{\nu}_{e}s$ and left-handed $\nu_{\mu}s$ (or
$\nu_{\tau}s$) on the interpretation of the neutrino signal from
supernova SN1987A. The distortion of the electron anti-neutrino
spectrum is of special interest since almost all the neutrinos
detected from SN1987A were $\bar{\nu}_{e}s$. I will also consider the
possibility of deriving an upper bound on the neutrino magnetic moment
from the observed neutrino spectra.

The conversion probabilities (i.e. the probability that a neutrino
emitted from the neutrinosphere has been converted into a different
species by the time it leaves the supernova) are taken from
\cite{ToSa}, which were calculated on the basis of the precollapse
model of Woosley and Weaver (1995) \cite{WW} for a progenitor mass of
15 and 25 solar masses and solar ($Z_\odot$) as well as zero ($Z_{0}$)
metallicity. The mass-squared difference $\Delta m^{2} =
m^{2}_{\nu_{\mu}} - m^{2}_{\bar{\nu}_{e}}$ and the vacuum mixing angle
$\theta_{\rm v}$ are assumed to take the values suggested by the
MSW-solution to the solar neutrino problem.

In the following section I will briefly review the theory of SF
oscillations. In section III I will discuss the various progenitor
models with their associated conversion probabilities and in section
IV I will present the numerical results which will be followed by the
discussion.

\section{Spin-Flavour Oscillations}

Assuming that the neutrino is a Majorana particle with a non-vanishing
transition magnetic moment, an anti-neutrino of one flavour can be
resonantly converted into a neutrino of another flavour on traversing
a magnetic field B and vice versa \cite{AkBe,LiMa,Vo}. These
spin-flavour oscillations are governed by the evolution equation

\begin{equation}
i\frac{d}{dr}\left ( 
\begin{array}{c} \bar{\nu}_{e} \\ \nu_{\mu}
\end{array}
\right ) = \left (
\begin{array}{cc} 0 & \mu_{\nu} B \\ \mu_{\nu} B 
& \Delta H 
\end{array}\right )\left ( 
\begin{array}{c} \bar{\nu}_{e}
\\ \nu_{\mu}
\end{array}\right ),
\end{equation}
where $\mu_{\nu}$ denotes the neutrino transition magnetic moment and
$B$ the component of the magnetic field which is transverse to the
direction of propagation of the neutrinos. The diagonal component of
the Hamiltonian is given by
\begin{equation}
\Delta H = \frac{\Delta m^{2}}{2E_{\nu}}\cos 2\theta_{\rm v} - \Delta V
\end{equation}
with $\Delta V = \sqrt{2}G_{\rm F}\rho /m_{\rm N}(1 - 2Y_{e})$, where
$\rho$ denotes the density, $G_{\rm F}$ the Fermi coupling constant,
$m_{\rm N}$ the nucleon mass, $E_{\nu}$ the neutrino energy and
$Y_{e}$ the electron fraction.  Resonance occurs when $\Delta H = 0$
and the conversion is adiabatic if

\begin{equation}
\gamma \equiv \frac{\mu B}{\sqrt{\left| \frac{d \Delta H}{d r}
\right|} } \geq 1 .
\end{equation}

If the neutrino is a Dirac particle, the effective potential is
proportional to $(1 - 3Y_{e})$, which is zero in the strongly
neutronized core. However the oscillations are highly non-adiabatic in
the high density core region for all but unrealistically high magnetic
fields so that a fine tuning would be required for any oscillations to
have a noticeable effect. Moreover SF-oscillations for Dirac neutrinos
are strongly constrained by observations \cite{Hi,Bi,No,LaCo} and
subsequently it will be assumed here that the neutrino is a Majorana
particle.

It has been pointed out by Athar et al. \cite{At} that in the
isotopically neutral (i.e. number of protons equals number of
neutrons) region above the iron core and below the hydrogen envelope
the effective matter potential is suppressed by 3-4 orders of
magnitude. Recently Totani and Sato \cite{ToSa} noted that the
potential is even further suppressed in the O+C and He layers in the
absence of heavy nuclei which, with their excess of neutrons, are
mostly responsible for the deviation from isotopic neutrality and it
turns out that the conversion probability is very sensitive to the
metallicity of the supernova progenitor.

According to the electroweak standard model the neutrino does not
possess a magnetic moment and an indication of a non-vanishing
magnetic moment would be a sign for physics beyond the standard model
and have profound implications for particle physics.

The current limit on the magnetic moment of the electron anti-neutrino
from laboratory experiments is $\mu_{\bar{\nu}_{e}} \leq 2.4 \times
10^{-10} \mu_{B}$ \cite{Vid,VoEn} for the Majorana neutrino.
Astrophysical constraints on the magnetic moment from the SN1987A
neutrino signal have been derived by N\"otzold \cite{No} and others
\cite{Pe,LaCo,RaWe} excluding values for $\mu_{\nu}$ of greater than a
few times $10^{-12}\mu_{B}$. For a review see \cite{PDG}. In the
following analysis a transition magnetic moment of around $\mu_{\nu}
\approx 10^{-12}\mu_{B}$ will be considered.

\section{Supernova Models and Conversion Probabilities}

The value of the effective matter potential $\Delta V$ was calculated
from the composition and density profiles of the precollapse model by
Woosley and Weaver (1995) \cite{WW}, which includes about 200 isotopes
up to $^{71}$Ge with solar abundances. The magnitude of $\Delta V$ as
function of radius is shown in Fig. 1.(a) and (b) for a progenitor
mass of 15 M$_\odot$ and 25 M$_\odot$ respectively. It is about
$10^{-5}$ eV in the centre and quickly decreases by about 10 orders of
magnitude as the density decreases and $Y_{e}$ approaches 0.5. Then at
around 0.1 R$_\odot$ the effective potential falls discontinuously by
several orders of magnitude and continues to decrease until it changes
sign at roughly the solar radius, where the He layer ends and the
hydrogen envelope starts. In this region $Y_{e}$ is very close to 0.5
(isotopically neutral) and the matter potential is suppressed by
several orders of magnitude.  In this region the deviation from
$Y_{e}=0.5$ is mainly due to isotopes like Ne, Mg, Al, S and
Ar. Totani and Sato \cite{ToSa} have reduced the abundance of these
isotopes over solar abundances thus suppressing the effective matter
potential even further. These models, subsequently called low
metallicity ($Z_{0}$) models, are shown in Fig. 1. (c) and (d). The
investigation of low metallicity progenitors seems especially
important given the low metallicity encountered in the LMC of around
$Z\approx Z_\odot /4$ and various other observations that have led to
the belief that the progenitor of SN1987A was a low metallicity blue
super-giant (BSG). For a study of the progenitor of SN1987A see
e.g. Langer {\em et al.} 1989 \cite{La}.  It can be seen from Fig.1
that in the $Z_{0}$ - models the isotopically neutral region is
broader and the effective potential in this region is suppressed by
another 1-2 magnitudes over the solar metallicity models.

For the purpose of calculating the conversion probabilities it can be
assumed that the composition of the envelope remains unchanged during
infall since the collapse is homologous and the shock wave cannot
reach the isotopically neutral region during the neutrino burst (1-10
s). Resonance occurs in the isotopically neutral region for values of
$10^{-10} \frac{{\rm eV}^{2}}{\rm MeV} < \frac{\Delta
m^{2}}{E_{\nu}}<10^{-1}\frac{{\rm eV}^{2}}{\rm MeV} $.

There is no direct information on the magnitude of magnetic fields
inside a collapsing star, but a rough estimate may be obtained from
the observed field on the surface of white dwarfs \cite{Cha}
suggesting magnetic fields of up to $B_{0}\approx 10^{10}$ Gauss at
the surface of the iron core. Assuming that the radial dependence of
the magnetic field above the core follows a dipole field, $B \propto
r^{-3}$, which would give a field in the isotopically neutral region
of about $10^{4}$ G. Higher estimates may be obtained by equating the
magnetic field energy to the thermal energy of the gas (see
\cite{ToSa}). In following analysis I will assume a dipole field with
field strengths at the surface of the iron core of $B_{0}\approx
10^{9}$ G and $B_{0}\approx 10^{10}$ G.

Also I assumed mixing parameters which are favoured by the
MSW-interpretation of the solar neutrino problem: the large angle
solution with $\Delta m^{2} \approx 10^{-6} {\rm eV}^{2}$,
$\sin^{2}2\theta_{v}\approx 0.7$, and the small angle solution with
$\Delta m^{2} \approx 6 \times 10^{-6} {\rm eV}^{2}$ and
$\sin^{2}2\theta_{v}\approx 10^{-2}$.

The conversion probabilities corresponding to the models discussed
above have been calculated by Totani and Sato \cite{ToSa} and some
typical cases are shown in Fig. 2 for $\Delta m^{2}=10^{-6} \: {\rm
eV}^{2}$ and $\sin^{2}2\theta_{v}\approx 10^{-2}$. For other mixing
angles or neutrino masses the energy can simply be rescaled with
$(\frac{{\rm eV}^{2}}{\rm MeV}\frac{E_{\nu}}{\Delta m^{2}}\cos
2\theta_{v})$.

These plots demonstrate that the magnitude of the conversion
probability as well as the dependence on energy are very different for
the solar and low metallicity models.  For solar metallicity the
conversion probability is generally very small for all but the most
extreme values of $B$ or very high values of $\Delta m^{2}$, which we
will not consider here. For a field strength of $B_{0}\approx 10^{9}$
G the conversion probability is negligible and not shown in Fig. 2 and
even for $B_{0}\approx 10^{9}$ G the conversion probability is hardly
significant. As expected the efficiency of the conversion improves
with increasing $\mu B$ (and thus increasing $\gamma$). The conversion
probability decreases with increasing $E_{\nu}$ as seen from the
dependence of $\gamma$ on $\frac{\Delta m^{2}}{E_{\nu}}$.  However for
progenitors of low metallicity the conversion probability increases
with energy and remains constant at high energies due to the
precession of the magnetic moment in the broad conversion region. The
conversion probability is periodic in $\mu B$ since the precession
length is dependent on $\mu B$, which determines the phase at the
outer edge of the isotopically neutral region. The conversion
probability for a low metallicity, 25 M$_\odot$ model is not shown
here. Like for the 15M$_\odot$ model conversion is effective for a
field of $B_{0}=10^{10}$ G but it contains a number of complicated
features due to precession in the inner part of the isotopically
neutral region.

 Also it is found that in contrast to the metallicity the dependence
 on the progenitor mass is comparatively weak \cite{ToSa}.

\section{Statistical Analysis}

In total 19 neutrinos from SN1987A have been detected: Kamiokande
recorded 11 and IMB eight events \cite{Hi,Bi}. A maximum likelihood
analysis is frequently used as an effective and unbiased method to
interpret this sparse data \cite{Je,Ja}.  I will be using the
likelihood function suggested by Jegerlehner {\em et al.}
\cite{Je}. If the expected spectrum of detected energies is $n(E)$,
the total number of events $N_{\rm obs}$ and the observed energies
$E_{i}$, the likelihood function ${\cal L}$ is given by
\begin{equation}
{\cal L} \propto \exp \left ( \int_{0}^{\infty}n(E)dE \right)\prod
^{N_{\rm obs}}_{i=1}n(E_{i}).
\end{equation}
For a joint analysis of the signal from both detectors, the likelihood
function is the product of the individual ${\cal L}$'s.  The expected
spectrum of detected energies is given by
\begin{equation}
n(E)=\int_{0}^{\infty}dE'P(E,E')\eta(E')n_{p}(E'),
\end{equation}
where $P$ denotes the probability of detecting an energy $E$ if the
real energy is $E'$, which is assumed to be Gaussian with an
energy-dependent dispersion:
\begin{equation}
P(E,E')=\left (\frac {2 \pi E_{s} E'}{\rm MeV^{2}}\right )^{-1/2}\exp
\left ( -\frac{(E-E')^{2}}{2E_{s} E'}\right ).
\end{equation}
where $E_{s}=0.75$ MeV for Kamiokande and $E_{s}=1.35$ MeV for IMB.
$\eta$ represents the efficiency curve for the respective
detector. Analytic fits to these curves \cite{Bu} for Kamiokande are
\begin{equation}
\eta(E) = 0.93-\exp[-(E/9.0 \:{\rm MeV})^{2.5}] 
\end{equation}
for $E \geq 7.0 \:{\rm MeV}$ and $\eta = 0$ for $E \le 7.0 \:{\rm
MeV}$.  For the IMB detector
\begin{equation}
\eta(E) = 0.3975\frac{E}{10 \:{\rm MeV}} - 0.02625\left(\frac{E}{10
\:{\rm MeV}}\right)^{2} - 0.59
\end{equation}
for $1.9 < E/10 \:{\rm MeV} < 7.6$, $\eta = 0.915$ for $E/10\: {\rm
MeV} > 7.6$ and $\eta = 0$ for $E/10 \:{\rm MeV} < 1.9$. Lastly
\begin{equation}
n_{p}(E)=\frac{N_{p}}{4\pi D^{2}}\sigma(E+Q)F_{\bar{\nu}_{e}}(E+Q),
\end{equation}
where $Q$ is the mass difference between neutron and proton, $Q=1.29$
MeV, $\sigma = 2.295 \times 10^{-44}\:{\rm cm}^{2}$ is the
cross-section for the superallowed reaction $\bar{\nu}_{e}(p,n)
e^{+}$, $F_{\bar{\nu}_{e}}$ is the electron antineutrino flux at the
detector, $D=50$ kpc is the distance to the supernova and $N_{p}$ is
the number of target protons in the detector: $N_{p}=1.43\times
10^{32}$ for Kamiokande and $N_{p}=4.55\times10^{32}$ for IMB.

The likelihood function is a function of the free parameters of the
model $x_{i}$ and the data $y_{i}$. The best-fit values
$\tilde{x_{i}}$ are the parameters which maximize the likelihood
function
\begin{equation}
{\cal L}({\bf \tilde{x},y})=max{\cal L}({\bf x,y}) 
\end{equation}
Under the assumption that ${\cal L}$ is Gaussian, the confidence
region around the best fit values $\tilde{x}_{i}$ can be estimated by
\begin{equation}
\ln {\cal L}({\bf \tilde{x},y}) - \ln {\cal L}({\bf x,y}) \leq
\frac{1}{2} \chi ,
\end{equation}
where $\chi$ is 2.3, 4.61 and 6.17 for the 68.3\%, 90\% and 95.4\%
confidence levels respectively.

In the presence of SF-oscillations $(\bar{\nu}_{e}\leftrightarrow
\nu_{\mu})$ the primary electron anti-neutrino spectrum becomes
distorted due to the admixture of the $\nu_{\mu}$ spectrum, so that
the flux at some distance from the source is related to the primary
spectra $F_{\nu}^{0}$ by
\begin{equation}
F_{\bar{\nu_{e}}}=(1-p)F_{\bar{\nu}_{e}}^{0}+pF_{\nu_{\mu}}^{0},
\end{equation} 
where the conversion probability p is a function of the neutrino
energy $E$ and the mixing parameters.  The time-integrated primary
spectra are approximated by a Maxwell-Boltzmann distribution
\begin{equation}
F_{\nu}^{0}(E)=C E^{2} e^{-E/T_{\nu}}
\end{equation}
with the parameters C and $T_{\nu}$. Although there is little reason
to believe that the actual spectra really follow a Maxwell-Boltzmann
distribution, it allows to fit the first and second moments, $<E>$ and
$<E^{2}>$, of the spectra and seems reasonable in the light of an
exponential cooling model.

Equipartition of energy between all (anti-) neutrino species is
 assumed, but different neutrino `temperatures' for the $\bar{\nu}_{e}
 s$ and $\nu_{\mu} s$, $T_{\bar{\nu}_{e}}$ and $T_{\nu_{\mu}}$, are
 allowed.  Due to their lower interaction cross-section the
 temperature of the $\nu_{\mu} s$ is higher and I define
\begin{equation}
T_{\nu_{\mu}}\equiv \alpha T_{\bar{\nu}_{e}}
\end{equation}
with $\alpha$ predicted to lie between 1.5 and 2.0.  The total energy
$E_{\rm t}$ is six times the energy emitted in $\bar{\nu}_{e} s$ which
is found to be
\begin{equation}
E_{t}= 36 CT^{4}.
\end{equation} 
As in Ref. \cite{Je} the detector background is expected to have a
negligible influence on the results and is ignored here.
 
\section{Numerical Results}

Using the method described above the best-fit values for
$T_{\bar{\nu}_{e}}$ and $E_{\rm t}$ were first found in the absence of
oscillations. The Kamiokande data yielded a value for
$T_{\bar{\nu}_{e}}$ of 2.6 MeV with a total energy of $E_{\rm t} = 4.8
\times 10^{53}$ erg while the IMB data gave $T_{\bar{\nu}_{e}} = 3.6$
MeV and $E_{\rm t} = 4.7 \times 10^{53}$ erg. These results are in
good agreement with the values found by Jegerlehner {\em et al.}
\cite{Je} of $T_{\bar{\nu}_{e}} = 2.5$ MeV and $E_{\rm t} = 4.9 \times
10^{53}$ erg for Kamiokande and $T_{\bar{\nu}_{e}} = 3.7$ MeV and
$E_{\rm t} = 5.4 \times 10^{53}$ erg for IMB and thus independently
confirm their results. 

Subsequently I calculated the best-fit parameters for a variety of
models.  Best fit values for $E_{\rm t}$ and $T_{\bar{\nu}_{e}}$
together with errorbars indicating the 90\% confidence levels are
shown in Fig. 3 and 4 for the Kamiokande and the IMB data
respectively.  For the $Z_\odot$ - models the mixing is too small to
have a significant effect and the results are virtually
indistinguishable from the ones obtained in the absence of
oscillations and not shown in Fig. 3 and 4. However the strong
conversion in the $Z_{0}$- models shifts the inferred
$T_{\bar{\nu}_{e}}$ to significantly lower values. Also a higher
relative temperature of the muon neutrinos (i.e. a higher $\alpha$)
lowers the inferred neutrino temperature. For the Kamiokande data
$T_{\bar{\nu}_{e}}$ is lowered to 1.9 MeV and 1.5 MeV for $\alpha=1.5$
and 2.0 respectively. Similarly for the IMB set the inferred value for
$T_{\bar{\nu}_{e}}$ is 2.5 MeV and 1.8 MeV.

The overlap between the values deduced from the two data sets is poor
and I refrain from a joint analysis of the data.

\section{Conclusion}

In agreement with other authors it is found that in the absence of
oscillations the results for the total energy released in neutrinos
are in excellent concordance with supernova theory (1), whereas the
temperature $T_{\bar{\nu}_{e}}$ lies slightly below the predicted
range (2). The lower end of the predicted range lies within the 90\%
confidence region.

As expected any admixture of the hotter $\nu_\mu$ spectrum through
SF-oscillations aggravates the discrepancy between the observed and
predicted neutrino spectra. It is seen that a bound on the neutrino
transition magnetic moment $\mu_{\nu}$ which is deduced from the
observation of SN neutrinos is strongly dependent on the metallicity
of the progenitor. It is found that a low metallicity can
substantially increase the sensitivity to $\mu$ and has to be taken
into account when deducing bounds on $\mu_{\nu}$ from supernova
neutrinos. Therefore bounds on $\mu_{\nu}$ derived from the neutrino
signal from SN1987A may be less stringent than earlier thought
\cite{At}. For solar metallicities the data from SN1987A is consistent
with a magnetic moment of $\mu_{\nu} \approx 10^{-12}\mu_{B}$ for any
realistic magnetic field inside the supernova and mixing parameters
chosen from the MSW solution to the solar neutrino puzzle. However
lower metallicities as found in the LMC make a magnetic moment as high
as $10^{-12}\mu_{B}$ seem less likely unless the magnetic field is
lower than argued here.

Thus under the assumption that the progenitor of SN1987A in fact had a
metallicity as low as $Z_{0}$ and a magnetic field as high as
$B_{0}\approx 10^{10}$ G the data seems to restrict the magnetic
moment to $\mu_{\nu} < 10^{-13}\mu_{B}$.

The aim of this article is to point out the dependence of the inferred
value for the neutrino magnetic moment on the metallicity of the
progenitor thus highlighting the uncertainties involved in
interpreting the data.  I should emphasize again that the poor
statistics of the data as well as the many unknowns of the source do
not allow a reliable determination of neutrino properties such as the
transition magnetic moment.  However there is hope that with the
advent of a new generation of neutrino telescopes as Superkamiokande,
the Large Volume Detector (LVD), the Sudbury Neutrino Observatory
(SNO) and others a future supernova explosions in our neighbourhood
will provide us with richer data allowing better estimates of neutrino
properties and variables of the supernova explosion.

\acknowledgements

I am grateful to Dr. T. Totani for providing me with the data for the
progenitor models. Also I thank the first referee for several useful
comments and Prof. W.C. Haxton for helpful discussions.

\begin{figure}
\caption{Magnitude of the effective potential $\Delta V$ as a function
of radius for several progenitor masses and metallicities. $\Delta V$
is measured in eV and the radius is measured in units of solar
radii. (a) 15 M$_\odot$, Z$_\odot$, (b) 15 M$_\odot$, Z$_{0}$, (c) 25
M$_\odot$, Z$_\odot$, (d) 25 M$_\odot$, Z$_{0}$ }
\label{fig1}
\end{figure}

\begin{figure}
\caption{Conversion probabilities as function of neutrino energy
$E_{\nu}$, solid line: Z$_\odot$, 15 M$_\odot$, $B_{0}=10^{10}$ G,
dotted line: Z$_\odot$, 25 M$_\odot$, $B_{0}=10^{10}$ G, dashed line:
Z$_{0}$, 15 M$_\odot$, $B_{0}=10^{10}$ G, dot-dashed line: Z$_{0}$, 25
M$_\odot$, $B_{0}=10^{9}$ G}
\label{fig2}
\end{figure}

\begin{figure}
\caption{Best fit values for $E_{\rm t}$ and $T_{\bar{\nu}_{e}}$ for
the Kamiokande data. The errorbars correspond to 90\% confidence
level.}
\label{fig3}
\end{figure}

\begin{figure}
\caption{The same as Fig. 3 but for the IMB data.}
\label{fig4}
\end{figure}

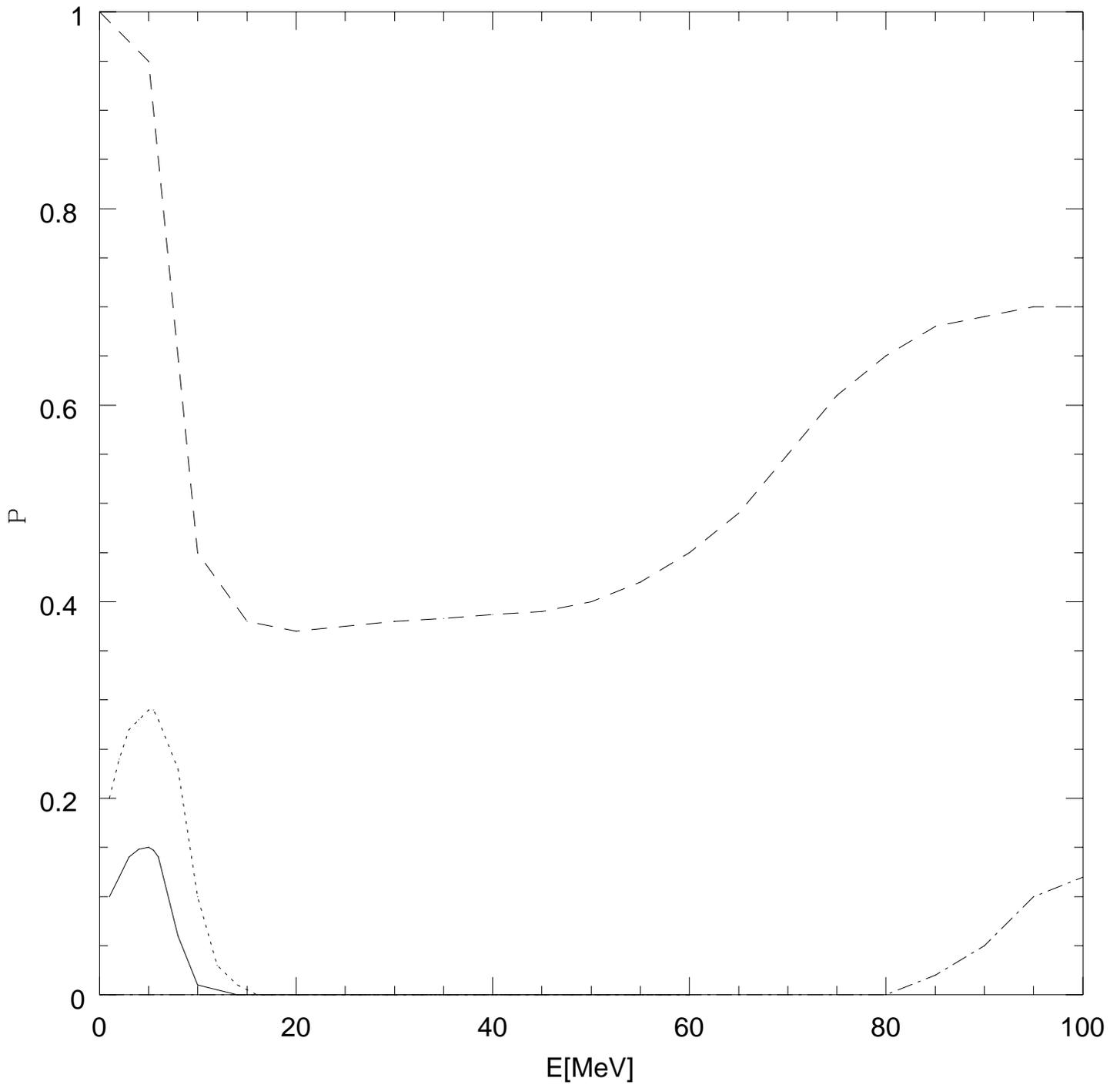
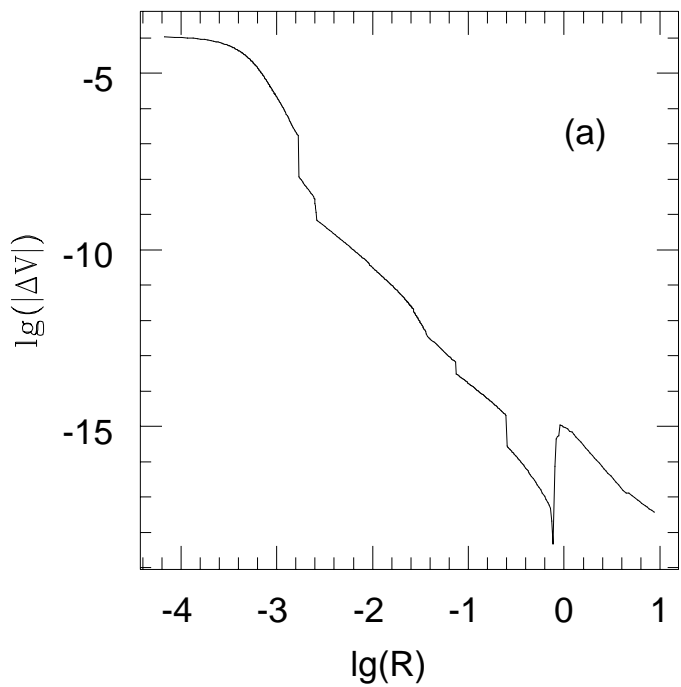
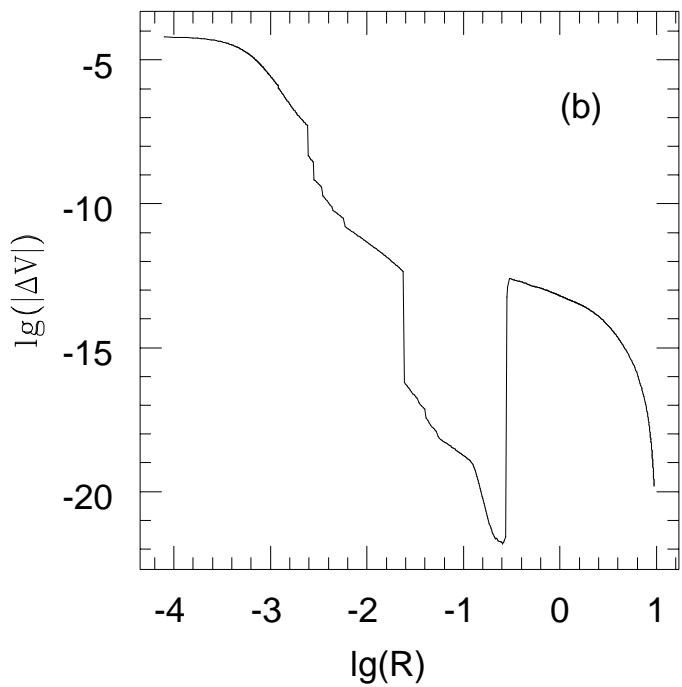
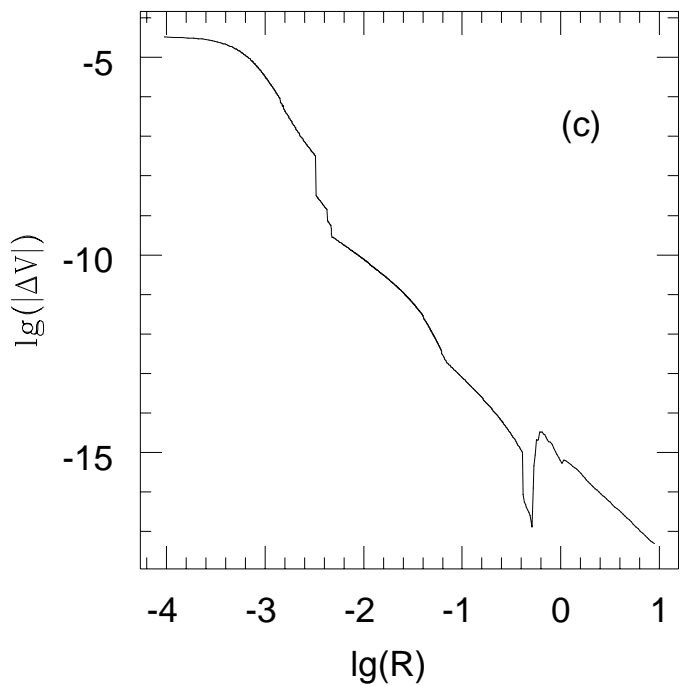
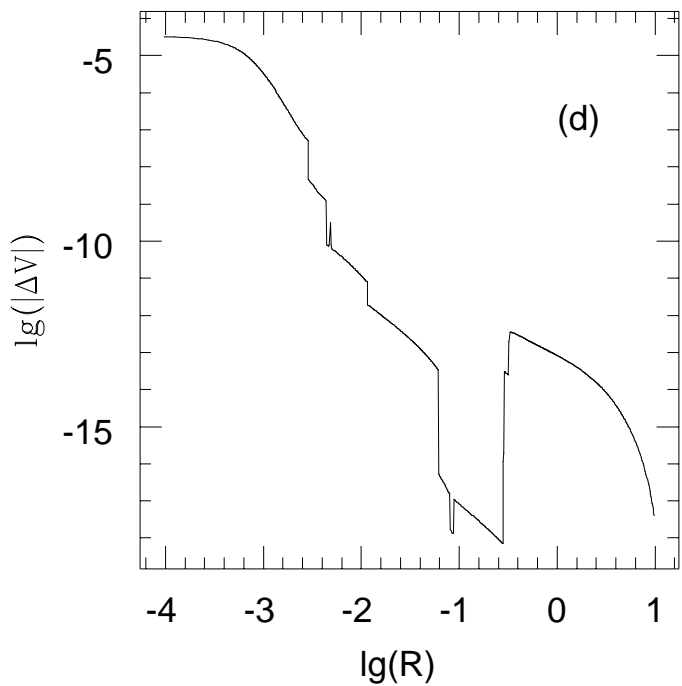
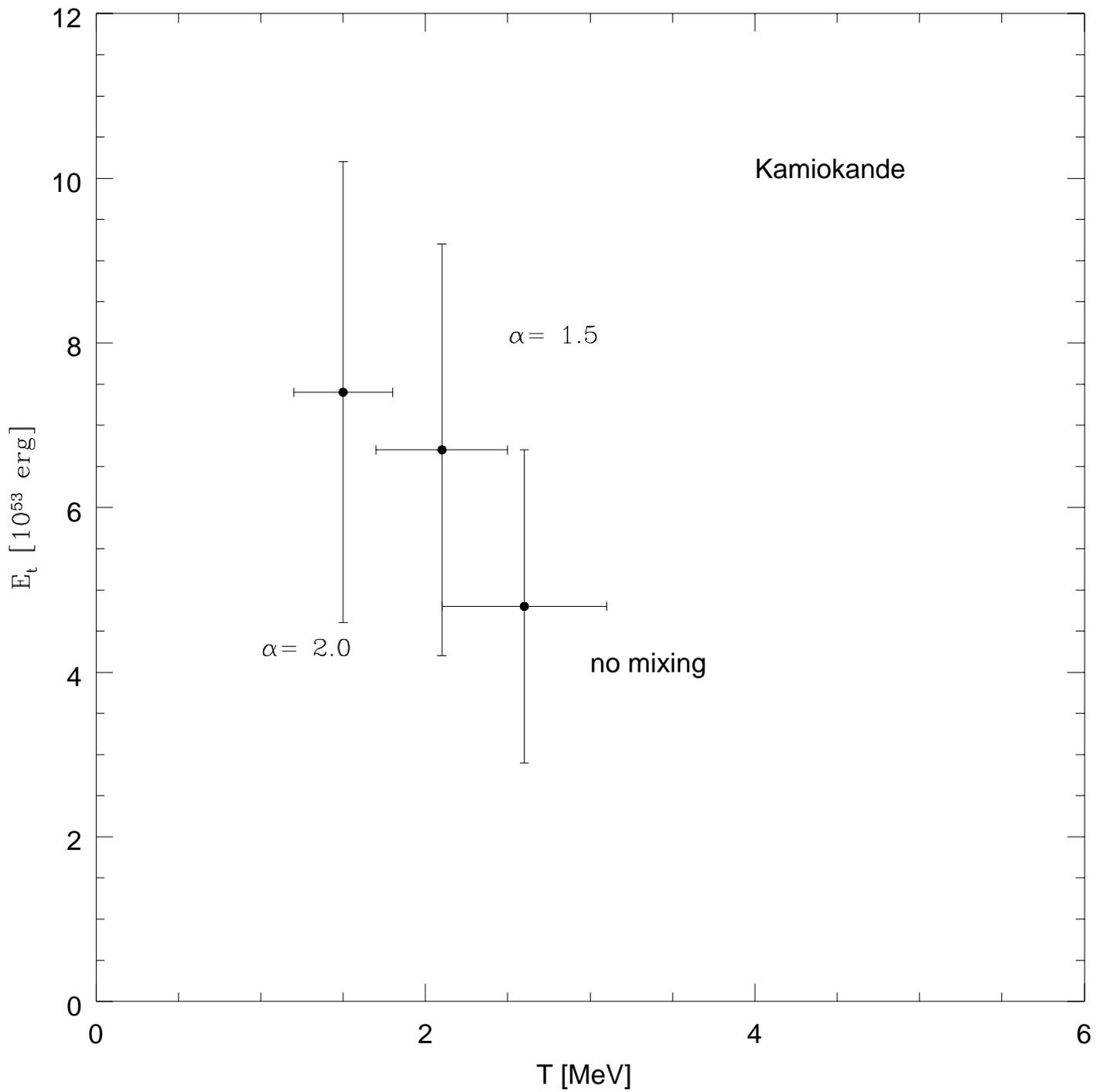
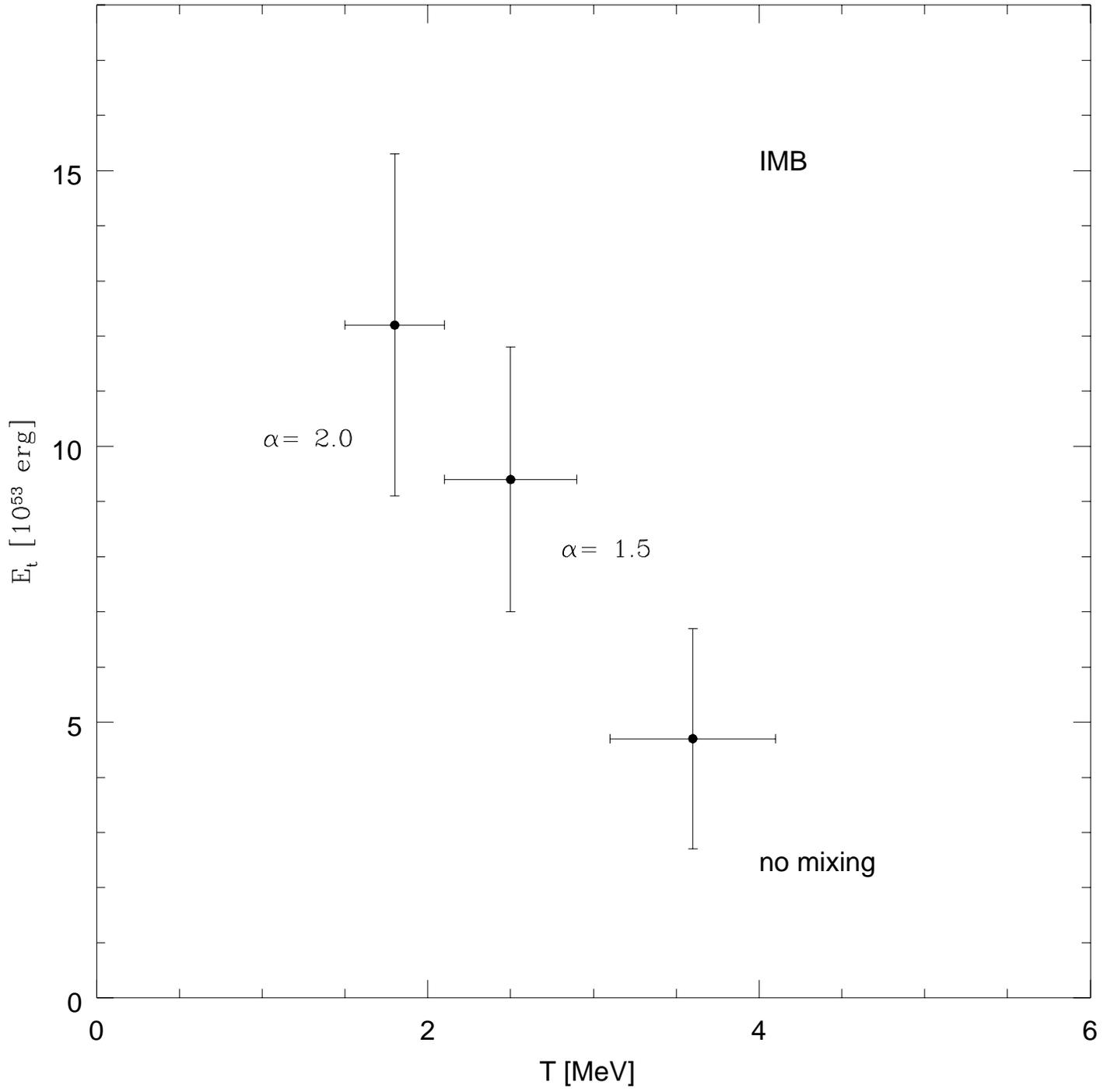
\end{document}